\documentclass[12pt]{article}
\usepackage{epsfig,epsf,rotating}
\begin{document}

\title{On a geometric mean and power-law statistical distributions.}

\author{A.Rostovtsev\\
\\
{\it Institute f. Theoretical and Experimental Physics, ITEP,} \\
{\it Moscow, Russia} \\
{rostov@itep.ru}
}
\date{ }
\maketitle
\begin{center}
Abstract
\end{center}
For a large class of statistical systems a 
geometric mean value of the observables is constrained.
These observables are characterized by a 
power-law statistical distribution.
\vspace*{1.0cm}

In everyday life we find events of very different nature often to follow 
similar statistical distributions.  These distributions arise because the 
same stochastic process is at work, and this process can be understood 
beyond the context of each example. Who is responsible for the particular 
forms of these distributions? This is a fundamental problem. The least 
bias approach to this problem was promoted by E.T.Jaynes about half 
century ago as a Maximum Entropy Principle (MEP)~\cite{Jaynes}. The MEP 
states that the physical observable has a distribution, consistent with 
given constraints which maximize the entropy. It is important that the 
{\it a priori}
given constraints are defined by a macroscopic mechanism the 
observable is constructed. If the observable values are restricted to be 
within a limited interval only, the resulting statistical distribution is 
a uniform distribution. A maximization of Shannon-Gibbs entropy with a 
constraint on the observable's arithmetic mean value results in an 
exponential distribution. Examples of the observables constructed such a 
way that an arithmetic mean values are constrained are following.
\begin{itemize}
\item[a)] Time until the 
next event, given $N$ events are uniformly distributed over a time period 
$T$. In this case 
the arithmetic mean value is constrained to be $T/N$.
\item[b)] energy of an ideal 
gas molecule has the arithmetic mean value constrained to be $E/M$, 
provided the isolated gas volume contains $M$ molecules with 
total energy $E$.
\item[c)] Time until a radioactive particle decays, with the arithmetic 
mean value given by the nature of the decaying particle. 
\end{itemize}
It is well known, the exponential distributions are observed for 
a)-c). Note, that the given examples haven't any other constraints than the 
arithmetic mean. An additional constraint would distort the exponential 
distribution form. Consider a set of unbiased measurements of a rod length 
by a ruler. The arithmetic mean value of these measurements is defined 
by true physical length of the rod and, in addition, the variance is 
defined by accuracy of the ruler. A maximization of Shannon-Gibbs 
entropy with constraints on the arithmetic mean and the variance results in 
the normal distribution, as it is usually observed for a set of repetitive 
measurements. Ubiquity of exponential distributions in the Nature implies 
that a vast class of processes has only one intrinsic constraint which is 
the arithmetic mean of a correspondent observable.
  
       However, the exponential distribution is not only ubiquitous one in 
the Nature. Power-law distributions occur in an extraordinary diverse 
range of phenomena in physics, biology, geography, economics, 
lexicography, computing, ecology, {\it etc}. The examples of observables 
having 
a power-law behavior are: a settlement population, people's annual 
incomes, the sizes of earthquakes, avalanches and landslides, the forest 
fires and solar flares, computer files, trades on a stock market, the 
occurrence of words in a book, the sales of commodities, a number of 
sexual contacts, particle energy in a geomagnetic plasma, hadrons produced 
in high energy particle interaction, {\it etc}~\cite{PL_reviews}. 
In order to obtain a power-law distribution 
from MEP one has to constrain an average logarithm of the observable 
value, or, equivalently, a geometric mean value. This seems to be odd 
since the use of geometric means is not generally anticipated. That is why 
it becomes increasingly attractive to derive the power-law distributions 
by exploiting MEP for non-extensive form of entropy with a 
traditional constraint on 
an arithmetic mean value~\cite{Tsallis}. We argue below that for a vast 
class of phenomena having a power-law statistics an arithmetic mean isn't 
defined from the first principles, while an application of a geometric 
mean constraint is rather logic. This has been already pointed out in 19th 
century by
Francis Galton who has stressed an importance of geometric means for 
vital 
and social statistics~\cite{Galton}.  Let us consider below some selected
examples in which geometric mean values get constrained. 

I. Population of settlements. Since 1913 it was observed that a 
distribution of the number of cities as function of it's human population 
follows a power-law~\cite{Auerbach}. A growth (decay) of cities is a 
complicated process which generally depends on political and economical 
situation, human migrations, birth (death) rate, {\it etc}. However, on a 
macroscopic level these factors are defined in terms of a relative rate of 
increasing (decreasing) population, i.e. a ratio of a city population 
change~$(\Delta{x})$ over certain time period~$(\Delta{t})$ to the city 
population~$(x)$. An 
average of this ratio could, in principle, be modeled {\it a priori}  
independently of  a city size. On the other hand the absolute value 
$\Delta{x}$ in 
not constrained and strongly depends on the city size~$x$.    
             
       II.         Frequency of occurrence of words in a book. Suppose 
that the words in a book are ranked according to their number of 
occurrences, with the 1st rank corresponding to the most frequent word.
In early 1930s, G.K.Zipf has noticed that for large ranks
the number of word occurrences has a power-law dependence as function of 
the word's rank~\cite{Zipf}. This 
remarkable observation holds true for different languages, authors and 
styles and is often termed as Zipf's law. A quantitative explanation of 
Zipf's law was later provided by H.A.Simon as a text generation 
model~\cite{Simon}. According to this model a probability to add a new or 
an already used word is a context dependent. The words that have been 
already used will be added to the text with a frequency~$(\Delta{n})$ per 
text 
unit, say line,~$(\Delta{l})$ proportional to the number of its previous 
occurrences. Thus, the model of the text generation defines a 
characteristic ratio of the word occurrence 
increment~$\Delta{n}/\Delta{l}$ 
over the prior occurrence~$n$.

III.	People's annual income. The distribution of people's income in all 
countries and cultures has a power-law form. This phenomenon is called 
Pareto law after the Italian economist Vilfredo Pareto, who has first 
reported it in the end of XIX century~\cite{Pareto}.  A toy model of 
wealth exchange within a society suggests that one invests 
each time (to gain or to loose) in average a certain fraction of one's 
income. After some iterations the number of poor people will rise up while 
a small fraction of rich people will build up a long tail of the 
distribution towards enormous richness. The poor get poorer and the rich 
get richer, but in case people are bind to invest in average a predefined 
fraction of their personal income the final distribution of people's 
income has a power-law form. Contrary to that, if people exchange a random 
amount of wealth, which is only limited by the size of income the final 
distribution of people's income is exponential~\cite{follow_money}. 

       IV.     Sexual contacts. The investigations of the nature of the 
web of human sexual contacts aimed at preventing the spread of 
sexually-transmitted diseases was undertaken in Sweden~\cite{sex}. The 
survey involving a random sample of 4781 Swedish individuals demonstrated 
that the distributions of the number of sexual partners during the twelve 
months prior to the survey decays as a power law with similar exponents 
for females and males. The models for building up the web of sexual 
contacts deal with an average frequency to take a decision to change the 
sexual partner. This frequency is different for different individuals and 
depends on the personal experience. The more partners $(n)$ one has 
changed 
in the past, the higher probability~$(dn/dt)$ to find a new partner in the 
future in average. Since a characteristic ratio $<dn/dt/n>$ could vary for 
different social groups, the reported observation was restricted to a 
certain age range with prostitutes excluded.
           
The above examples (I-IV) have a common feature: the evolution of the 
physical observables is a statistical process which is determined by a 
probability~$(P)$ as function of a relative change of the observable per 
an evolution step. This distribution has a characteristic scale of 
the evolution defined by
\begin{equation}
\langle \frac{1}{x}\frac{dx}{dt}\rangle = 
P \otimes
\Bigl(\frac{1}{x}\frac{dx}{dt}\Bigr) = a\,.
\label{P}
\end{equation}
Therefore, for a statistical ensemble of $N$ observable values~$(x_i)$
\begin{equation}
\frac{1}{N}\sum_{i=1}^N \frac{1}{x_i}\frac{dx_i}{dt} = 
\frac{1}{N}\sum_{i=1}^N \frac{d}{dt}log(x_i) = a\,.
\end{equation}
After adding up many subsequent evolution steps one finds
\begin{equation}
\frac{1}{N}\sum_{i=1}^N log(x_i) = A + a\cdot T\,,
\end{equation} 
and
\begin{equation}
\Bigl(\prod_{i=1}^N x_i \Bigr)^\frac{1}{N} = exp(A + a\cdot T)\,.
\end{equation}
Here, for simplicity, the parameter $a$ is assumed to vary slowly with 
time. 
In fact, if at the end of the evolution the final probability distribution 
becomes stable, the parameter $a$ turns to zero as it is expected for a 
system at equilibrium.
  Thus, for the discussed class of the statistical systems a geometric 
mean is constrained 
by the system's evolution prescriptions, whilst an arithmetic mean or, 
equivalently, a characteristic scale of $<dx/dt>$,  has a little 
sense and could be only deduced from the measurements. From this point of 
view, the attempts to derive the power-law statistics for~I-IV 
by exploiting MEP 
for non-extensive form of entropy with a constraint on an arithmetic mean 
value would look unjustified. 
  Note, if the final distribution of the observable $x$ follows exactly a 
power-law, then the geometric mean is the only one constraint 
which is available. 
If so, the MEP suggests the probability distribution~$P$ in~(\ref{P}) to 
be exponential. 
Indeed, it is likely, for the words with a given frequency of occurrence 
in a book a 
text length until the word occurs next time has an exponential 
distribution. Similarly, for a particular category of individuals who 
change a given number of sexual partners per year a waiting time until the 
next partner is also likely to have an exponential distribution.  Thus, 
for the discussed above examples (I-IV) one could generate the $dx/dt$ 
distribution using an exponential distribution~$(\epsilon)$ 
\begin{equation}
\frac{dx}{dt} = x\cdot \epsilon\,.
\end{equation}
Contrary to that, for a class 
of observables having an exponential distribution 
\begin{equation} 
\frac{dx}{dt} = b\cdot \epsilon\,,
\end{equation}
with a constant model parameter $b$. In this case the evolution of the 
system is determined by an absolute change of the observable per an 
evolution step. This evolution mode is realized, {\it e.g.}, 
in the radioactive atom decays, where each time the decay 
probability is independent of how long the atom survived before. There is 
no good reason to forbid a mixed evolution model with
 \begin{equation} 
\frac{dx}{dt} = x\cdot \epsilon + b\cdot \epsilon\,. 
\end{equation}
For such mixed model the arithmetic mean value is
\begin{equation}
\langle \frac{1}{b+x}\frac{dx}{dt}\rangle = a\,,
\end{equation}
and, therefore, a corresponding constraint on $x$ is defined as
\begin{equation}
\langle log(b+x) \rangle = A+a\cdot T\,.
\label{log}
\end{equation}
A maximization of Shannon-Gibbs entropy with the 
constraint~(\ref{log}) 
results in a damped power-law probability distribution 
\begin{equation}
\frac{1}{(b+x)^\alpha}\rightarrow 
{\biggl \{} {{exp(-x/\alpha{b}), x\rightarrow 0} \atop 
{\frac{1}{x^\alpha},\,\,\,\,\,\,\,\,\,\,\,\,\,\,\,\,\,\,\,\, x\rightarrow 
\infty}}
\label{plaw}
\end{equation}
For small values of $x$ the damped power-law function~(\ref{plaw}) is 
approximated by 
exponential distribution, while for large values of $x$ it has a power-law 
behavior. Contrary to the power-law distribution the damped power-law 
function is not divergent at low $x$ values. This mixed behavior seems to 
be favorable in the nature. Many observables which initially are 
found to exhibit a power-law behavior later, after the accurate 
measurements 
at small $x$ values have been performed, turned out to follow very closely 
a damped 
power-law distribution. So, in~\cite{Zanette} the data on the word 
frequency 
occurrence are fitted to the damped power-law function. This implies 
a correction to the mentioned 
above toy model of text generation. It is plausible that the
most frequent "service" words~(like "a", "the", {\it etc}) 
have to be added to the text with a constant probability 
independent of the context of the book.  Among other numerous examples of 
the observables which are amazingly well  follow a
damped power-law distribution are trade sizes~\cite{trade_size} and 
inter-trade time intervals~\cite{Leonidov} on stock markets, hadron 
production in high energy particle collisions~\cite{HSQCD} and particle 
energy in a geomagnetic plasma~\cite{geomagnetic}. For the latter the 
damped power-law distribution is termed a "kappa-distribution".

     Finally, there are events of vital importance found to follow a 
power-law or damped power-law statistics. A mechanism of the most these 
phenomena is still poorly understood. 
Those phenomena are the trades at stock markets~\cite{trade_size, 
Leonidov}, global terrorism~\cite{terror}, extinction of biological 
species~\cite{extinction}, {\it etc}. For these and many other events with 
empirically known statistical distributions the MEP provides important 
information about the constraints at macroscopic level and, therefore, is 
helpful to understand the intrinsic mechanism of the phenomena.   

The author would like to thank A.Kropivnitskaya and T.Tokareva for helpful 
discussions. This work was initially motivated by a study of hadronic 
spectra and was partially supported by
Russian Foundation for Basic Research, grants 
RFBR-03-02-17291 and RFBR-04-02-16445.


\begin{thebibliography}{99}
\bibitem{Jaynes}
E.T.Jaynes, Information theory and statistical mechanics, Phys.Rev. 
{\bf 106} (1957) 620.

\bibitem{PL_reviews} 
D.Sornette, Critical phenomena in natural science, Springer, Heidelberg, 
(2003).\\
M.Mitzenmacher, A brief history of generative models 
for power law and lognormal distributions, Internet Mathematics, {\bf 1} 
(2004) 226.

\bibitem{Tsallis}
C.Tsallis, J.Stat.Phys., {\bf 52}, 479, 1988.\\
C.Tsallis, Brazilian J. of Rhys., {\bf 29}, 1, 1999.

\bibitem{Galton} 
F.Galton, The geometric mean in vital and social statistics, Proc.R.Soc. 
{\bf 29} (1879) 365.


\bibitem{Auerbach}
F.Auerbach, Das Gesetz der Bevolkerungskonzentration, Petermanns 
Geographische Mitteilungen, {\bf 59} (1913) 74.

\bibitem{Zipf}
G.K.Zipf, The psycho-biology of language, Houghton Miffin, Boston (1935).

\bibitem{Simon}
H.A.Simon, On a class of skew distribution functions, Biometrika, {\bf 42} 
(1955) 425.

\bibitem{Pareto}
V.Pareto, Cours d'Economie politique, Droz, Geneva (1896).

\bibitem{follow_money}
B.Hayes, Follow the money, American Scientist, {\bf 90} (2002) 400.

\bibitem{sex}
F.Liljeros, {\it et al}, The web of human sexual contacts, 
Nature, {\bf 411} (2001) 907.

\bibitem{Zanette}
D.H.Zanette, Zip's law and the creation of musical context, 
arXiv:cs.CL/0406015

\bibitem{trade_size}
V.Plerou, {\it et al}, On the origin of power-law fluctuations in stock 
prices, Quantitative Finance, {\bf 4} (2004) C11.

\bibitem{Leonidov}
A.Leonidov, Long memory in stock Trading, arXiv:cond-mat/0303222.

\bibitem{HSQCD}
A.Kropivnitskaya, A.Rostovtsev, Universalities in hadron production and 
the maximum entropy 
principle, Contribution to the HSQCD 2004 Conference, St.Petersburg 
(2004).'
 
\bibitem{geomagnetic}
M.A.Leubner, A Non-extensive Entropy Approach to Kappa-Distributions, 
Astrophysics and Space Science, {\bf 282} (2002) 573. 

\bibitem{terror} 
A.Clauset, M.Young, Scale invariance in global terrorism, 
arXiv.org/abs/physics/0502014.

\bibitem{extinction}
R.V.Sole and S.C.Manrubia, Extinction and self-organized criticality in a 
model of large-scale evolution, Phys.Rev.E, {\bf 54} (1996) R42.

\end{thebibliography}
\end{document}